\pdfoutput=1 
\documentclass[10pt,aps,prb,twocolumn,superscriptaddress,floatfix,nofootinbib]{revtex4-2}

\usepackage[utf8]{inputenc}
\usepackage[english]{babel}
\usepackage[T1]{fontenc}
\usepackage{physics} 
\usepackage{csquotes}
\usepackage{graphicx}

\usepackage{amsmath,amsfonts}

\usepackage[colorlinks=true]{hyperref}
\usepackage[capitalise]{cleveref}
\usepackage{xspace}
\usepackage{listings}

\usepackage{bm} 
\usepackage{xcolor}
\usepackage[normalem]{ulem}
\usepackage{orcidlink}

\renewcommand{\vec}[1]{\mathbf{#1}}
\newcommand{\iden}{\mathbb{I}}
\newcommand{\Esh}{E_{\textrm{shift}}}
\newcommand{\dt}{\delta t}
\newcommand{\dtau}{\delta \tau}
\newcommand{\tV}{$t$-$V$\xspace}
\newcommand{\neuralmodelname}{BiViT}
\newcommand{\mcest}[2]{\mathbb{E}_{#1}\left[#2\right]}
\newcommand{\Pf}{\operatorname{Pf}}
\newcommand{\Var}[1]{\operatorname{Var}\!\left[#1\right]}

\newcommand{\orcidjannes}{\orcidlink{0000-0001-7491-3660}}
\newcommand{\orcidjuan}{\orcidlink{0000-0001-7263-3462}}
\newcommand{\ethzurich}{Institute for Theoretical Physics, ETH Z\"urich, 8093 Z\"urich, Switzerland}

\begin{document}

\title{Fermionic neural Gibbs states}
\author{Jannes Nys$^{\orcidjannes}$}
\email{jannys@ethz.ch}
\affiliation{\ethzurich}
\author{Juan  Carrasquilla$^{\orcidjuan}$}
\email{jcarrasquill@ethz.ch}
\affiliation{\ethzurich}

\begin{abstract}

We introduce fermionic neural Gibbs states (fNGS), a variational framework for modeling finite-temperature properties of strongly interacting fermions. fNGS starts from a reference mean-field thermofield-double state and uses neural-network transformations together with imaginary-time evolution to systematically build strong correlations. Applied to the doped Fermi–Hubbard model, a minimal lattice model capturing essential features of strong electronic correlations, fNGS accurately reproduces thermal energies over a broad range of temperatures, interaction strengths, even at large dopings, for system sizes beyond the reach of exact methods. These results demonstrate a scalable route to studying finite-temperature properties of strongly correlated fermionic systems beyond one dimension with neural-network representations of quantum states. 
\end{abstract}

\maketitle
Accurately simulating thermal (Gibbs) states of quantum many-body systems is notoriously difficult, yet indispensable for elucidating phases of matter, thermodynamic response, transport, and nonequilibrium relaxation~\cite{eisert2015quantum,gogolin2016equilibration}. Thermal ensembles describe most laboratory conditions and provide the natural language to compare theory, numerical data, and experiments on an equal footing. However, numerical methods become particularly demanding when dealing with fermionic degrees of freedom.

Several families of classical finite-temperature algorithms have been developed. Among these, tensor-network approaches ---Matrix Product Operators (MPOs) and Projected Entangled-Pair Operators (PEPOs)--- approximate Gibbs states by exploiting locality and area-law structure~\cite{verstraete2004, zwolak2004, kliesch2014, molnar2015, kshetrimayum2019, vanhecke2023simulatingthermaldensity}. Typicality-based methods, including Thermal Pure Quantum (TPQ) states~\cite{sugiura2012,sugiura2013} and Minimally Entangled Typical Thermal States (METTS)~\cite{white2009,stoudenmire2010,bruognolo2017,takai2016finite}, reduce thermal averages to ensembles of pure-state simulations. Recent work targets the low-temperature regime via low-rank constructions seeded from zero temperature~\cite{cocchiarella2025low}. While originally introduced for spin degrees of freedom, these methods can typically be extended to predict thermal properties of fermionic systems as well, but tend to be limited to low-entanglement and low-dimensional systems.
In recent years, quantum Monte Carlo (QMC) methods have pushed the frontier of fermionic simulation~\cite{barker1979quantum,simkovic2019determinant,shen2020finite,rossi2017determinant,simkovic2020cdet,spada2021high,iskakov2024perturbative,vsimkovic2024origin,vsimkovic2022two}. Nevertheless, at strong doping, in frustrated lattices, and/or at low temperatures, the severity of the sign problem restricts accessible system sizes and parameter regimes.

Complementing purely numerical approaches, fermionic analog quantum simulators directly realize many-body Hamiltonians in the laboratory. Ultracold Fermi gases and fermionic atoms in optical lattices now routinely explore thermal regimes with strong correlations in more than one spatial dimension, reaching regimes at the edge of classical numerical feasibility, and beyond~\cite{bloch2008many,esslinger2010fermi,schneider2012fermionic, hart2015observation,koepsell2019imaging,hartke2023direct, xu2023frustration, prichard2024directly, lebrat2024observation, fabritius2024irreversible, bourgund2025formation}. 

Neural-network quantum states (NQS)~\cite{carleo2017} have emerged as powerful variational models for ground and low-energy excited states and the quantum dynamics of correlated quantum many-body systems~\cite{wu2023variational,schmitt2020dynamics,gutierrez2022real,donatella2022,sinibaldi2023unbiasing,medvidovic2022rotordynamics,nys2024ab, moss2025leveraging, hibatallah2020recurrent, pescia2024message, romero2025spectroscopy, moreno2022fermionic, viteritti2023transformer, pfau2020ab, pfau2024accurate, hermann2020deep}.
Extending these ideas to open systems for spin systems has been explored in Refs.~\cite{torlai2018latent, vicentini2019,nagy2019,hartmann2019,yoshioka2019, nomura2021purifying, carrasquilla2019,carrasquilla2021,luo2022, reh2021time, vicentini2022}, and in particular for finite temperature states in  Refs.~\cite{irikura2020finite,nomura2021purifying,nys2024real, kumar2025autoregressive, hendry2022neural}. More recently, the thermofield double formalism allowed Ref.~\cite{nys2024real} to demonstrate the efficacy of NQS in simulating the real-time dynamics of correlated thermal spin states in two dimensions. Despite impactful progress in modeling fermionic eigenstates~\cite{moreno2022fermionic, romero2025spectroscopy, chen2025neural, gu2025solving, kim2024ultracold, hermann2020deep, pfau2020ab, pfau2024accurate} and dynamical properties~\cite{nys2024ab}, a reliable and scalable approach to fermionic Gibbs states is lacking.

We introduce fermionic neural density operators (fNDO) and, as a subclass, fermionic neural Gibbs states (fNGS): variational representations of fermionic thermal ensembles that (i) incorporate exchange antisymmetry by construction, (ii) capture fermionic Gibbs ensembles, and (iii) remain scalable across dimensions, temperatures, and interaction strengths. We demonstrate that fNGS accurately reproduces thermodynamics and correlations in benchmark lattice models and experimentally relevant regimes, and can represent strongly correlated fermionic density matrices at finite temperature.

\section{Method}
We generate fermionic thermal states through the thermofield double (TFD) formalism~\cite{harsha2020wave, harsha2023thermofield, henderson2020correlating, nys2024real}. In the TFD approach, thermal properties of a system are computed via a single imaginary-time evolution of an entangled purification of the thermal density matrix.
We represent this state by a variational ansatz constructed by supplementing well-understood mean-field states with neural-network transformations~\cite{luo2019backflow, kim2024ultracold, lou2024neural, pescia2024message} acting on two identical copies of the system. 

\subsection{Mean-field thermofield doubled fermions}
The starting ingredient in our approach is the construction of a finite-temperature mean-field reference state, which also lays the groundwork for extending the formalism to interacting fermions in the next subsection. Consider a mean-field Hamiltonian $\hat{H}_0$ with orthonormal eigenstates $\ket{\mu}\equiv \hat{c}_\mu^\dagger \ket{\Omega}$ with energies $\epsilon_\mu$ and vacuum $\ket{\Omega}$. In the TFD formalism, we introduce a double Hilbert space $\mathcal{H} \otimes \tilde{\mathcal{H}}$. Fermionic operators in the $\tilde{\mathcal{H}}$ Hilbert space are denoted $\tilde{c}_\mu^\dagger$ where we use $\hat{\phantom{o}}$ and $\tilde{\phantom{o}}$ to denote operators in the physical and auxiliary thermofield space, respectively. We impose the anti-commutation relations $\{\hat c_\mu^{(\dagger)},\tilde c_\nu^{(\dagger)}\}=0$. A thermal state at inverse temperature $\beta=1/T$ is given by 
\begin{align}
    \ket{\Psi(\beta)} 
    &= 
    \prod_\mu 
    \left[
        u_\mu(\beta)
        + 
        v_\mu(\beta) 
        \hat{c}_\mu^\dagger \tilde{c}_\mu^\dagger
    \right]
    \ket{\Omega}.\label{eq:psi_beta_mf}
\end{align}
With $w_\mu(\beta) = e^{-\beta \epsilon_\mu / 2}$ we have $u_\mu(\beta) = 1/\sqrt{1 + w_\mu(\beta)^2}$, and $v_\mu(\beta) = w_\mu(\beta) u_\mu(\beta)$, where the latter yields the Fermi-Dirac occupation $|v_\mu|^2$.

This form resembles a Bardeen-Cooper-Schrieffer (BCS) state of physical-auxiliary fermion pairs, where $P_\mu^\dagger = \hat{c}_\mu^\dagger \tilde{c}_\mu^\dagger$ is a geminal creation operator and creates pairs of physical- and auxiliary-mode fermions. The finite-temperature expectation value of an operator $\hat{O}_\beta$ can be obtained as (see also Ref.~\cite{nys2024real} and Appendix~\ref{sec:observables} for Monte Carlo estimators)
\begin{align}
    \expval{\hat{O}}_\beta &= \frac{\mel{\Psi(\beta)}{\hat{O} \otimes \tilde{\iden}}{\Psi(\beta)}}{\braket{\Psi(\beta)}}
\end{align}
 
We consider systems in the canonical ensemble, and hence projecting both systems in Eq.~\eqref{eq:psi_beta_mf} onto a fixed charge sector with $N$ fermions,  yields~\cite{harsha2020wave} (up to normalization from hereon)
\begin{align}
\mathcal{P}_N \ket{\Psi(\beta)} 
&= [\Gamma^\dagger(\beta)]^N \ket{\Omega} = \ket{\mathrm{AGP}}
\end{align}
with $\Gamma^\dagger(\beta) = \sum_\mu w_\mu(\beta) P_\mu^\dagger$ and $\ket{\mathrm{AGP}}$, an antisymmetric geminal power state~\cite{khamoshi2020correlating, lou2024neural}, see Appendix~\ref{sec:pfaffian}. 

Finally, we consider the projection of this mean-field thermal state onto a specific Fock state. In particular, consider a fermion pair in mode $p$ (with creation operator $\hat{f}^\dagger_p$) and thermofield mode $q$ (i.e.\ $\ket{p, \tilde{q}} = \hat{f}_p^\dagger \tilde{f}_q^\dagger \ket{\Omega}$), 
\begin{align}
    \mel{p, \tilde{q}}{\Gamma^\dagger(\beta)}{\Omega}
    &= \sum_\mu w_\mu(\beta) \mel{p}{\hat{c}_\mu^\dagger}{\Omega} \mel{\tilde{q}}{\tilde{c}_\mu^\dagger}{\Omega}  \\
    &= \sum_\mu w_\mu(\beta) \phi_\mu(p) \tilde{\phi}_\mu(q) \\
    &=: \varphi(p, q) \label{eq:pair_varphi}
\end{align}
where $\phi_\mu(p) = \mel{p}{\hat{c}_\mu^\dagger}{\Omega}$, and we expressed the systems in a conjugated orbital basis $\tilde{\phi}_\mu = \phi_\mu^*$. 
A many-body thermal mean-field state for a Fock state $\ket{x, \tilde{y}}$ occupying modes $(p_1, ..., p_N, \tilde{q}_1, ..., \tilde{q}_N)$ is fully parametrized by the temperature-dependent pair matrix $[\varphi]_{p,q} := \varphi(p,\tilde q)$., 
\begin{align}
    \braket{x, \tilde{y}}{\Psi(\beta)} &= \sum_{\mathcal{S}} e^{-\beta E(\mathcal{S})/2} \braket{x}{\mathcal{S}} \braket{\tilde{y}}{\mathcal{S}} \\
    &= \det \begin{pmatrix}
         [\varphi]_{p_1, \tilde{q}_1} & \dots  & [\varphi]_{p_1, \tilde{q}_N} \\
        \vdots & \ddots & \vdots\\
        [\varphi]_{p_N, \tilde{q}_1} & \dots  & [\varphi]_{p_N, \tilde{q}_N} 
    \end{pmatrix} \label{eq:det_pair}
\end{align}
Here, $p=(l, \sigma)$ will typically refer to an occupied mode with spin $\sigma$ and orbital (or site) $l$. For spin-conserving Hamiltonians, the determinant factorizes into two spin contributions.
The set $\mathcal{S}$ runs over all combinations of $N$ occupied states, and we used the Cauchy-Binet identity~\cite{broida1989linear} to obtain the thermal pair orbitals in Eq.~\eqref{eq:pair_varphi}.

\subsection{Correlated thermofield fermions}
The analytic mean-field expression of the thermal state in Eq.~\eqref{eq:det_pair} serves as the foundation for incorporating many-body correlations and constructing an ansatz for correlated fermionic thermal states. We define our correlated fermionic neural Gibbs states (fNGS) by first building correlations into the mean-field state at a fixed temperature. We generalize the mean-field formalism by lifting the pair orbitals to matrix functions that depend on the many-body configurations $[\varphi(x, \tilde{y})]_{p,q}$. We will focus on a correlated fNGS ansatz inspired by neural backflow~\cite{luo2019backflow} (yielding a second quantized form of the neural Pfaffian~\cite{kim2024ultracold} or neural AGP~\cite{lou2024neural} used to capture pairing in superfluids)
\begin{align}
    &\braket{x, \tilde{y}}{\Psi(\beta)} = \nonumber \\
    &\det \begin{pmatrix}
        [\varphi(x, \tilde{y})]_{p_1, \tilde{q}_1} & \dots  & [\varphi(x, \tilde{y})]_{p_1, \tilde{q}_N}\\
        \vdots & \ddots & \vdots\\
        [\varphi(x, \tilde{y})]_{p_N, \tilde{q}_1} & \dots  & [\varphi(x, \tilde{y})]_{p_N, \tilde{q}_N}] 
    \end{pmatrix} e^{-J(x, \tilde{y})}
\end{align}
where $\varphi$ is a neural network, and $J$ is a Jastrow-like factor capturing both inter- and intra-species interactions. The details of the model with inter and intra-attention mechanisms are detailed in Appendix~\ref{sec:model}, and we discuss alternative ansatze in Appendix~\ref{sec:alternative_ansatz}. We will refer to our neural architecture as the \neuralmodelname, (an adaptation of the ViT~\cite{viteritti2023transformer}). Crucially, our model generates neural representations that capture both inter- and intra-species correlations. 
Alternative strategies to the neural backflow transformations~\cite{luo2019backflow} are possible, such as hidden fermion pairs, a generalization of Ref.~\cite{moreno2022fermionic}.

\subsection{Variational imaginary time evolution}
Variational methods such as tensor networks often prepare Gibbs states via imaginary-time evolution from an infinite-temperature purification~\cite{verstraete2004, zwolak2004, kliesch2014}. Although most effective at high temperatures, this approach suffers from accumulated numerical errors when cooling toward low temperatures, and several refinements have been proposed to mitigate these issues~\cite{molnar2015, kshetrimayum2019, vanhecke2023simulatingthermaldensity}. Similar difficulties arise in variational Monte Carlo, where direct cooling from infinite temperature remains challenging even in spin systems~\cite{nys2024real}.
Here, we adopt a different strategy inspired by Ref.~\cite{holmes2022quantum}. Instead of cooling from infinite temperature, we begin from a thermal state of a solvable mean-field Hamiltonian $\hat{H}_0$ at inverse temperature $\beta_0$, characterized by geminal orbitals $\varphi_0$. From this reference state $\ket{\Psi_0(\beta_0)}$ in Eq.~\eqref{eq:det_pair}, we reach the correlated thermal state of the target Hamiltonian $\hat{H}$ by imaginary-time evolution with the work operator
\begin{align}
W &= \hat{H} \otimes \tilde{\iden} - \tfrac{\beta_0}{\beta} \hat{\iden} \otimes \tilde{H}_0 , \\
\ket{\Psi(\beta)} &= e^{-\frac{\beta}{2} W}\ket{\Psi_0(\beta_0)} .
\end{align}
Unless stated otherwise, we take $\beta = \beta_0$, with further cooling achievable via evolution under $\hat{H}\otimes\tilde{\iden}$. The resulting TFD state is symmetric under exchange of the physical and auxiliary systems, a property that can be imposed on the fNGS ansatz (see Appendix~\ref{sec:model}).

To implement this procedure efficiently, we develop a Taylor-root expansion projected imaginary-time evolution (tre-pITE) method, a high-order scheme for the imaginary-time propagator based on complex-time detours, following Ref.~\cite{nys2024ab}. When applicable, accelerate the projected imaginary‑time evolution by choosing each time step $\dtau$ adaptively: $\dtau=\sqrt{I_{c}/\Var{\smash{\hat H}}}$ (see Appendix~\ref{sec:adaptive_trepITE}). In addition, we propagate the variational parameters with a momentum‑like predictor across time steps, extrapolating the velocity extracted from the two most recent time-step solutions, which gives a reliable initial guess for the next optimization. Full details of the scheme are provided in Appendix~\ref{sec:adaptive_trepITE}.

Few other works have attempted to model electronic thermal states with fermionic NQS. As a first example, Ref.~\cite{xie2023m} proposed a variational density-matrix approach based on a neural canonical transformation to study the finite-temperature electron gas in continuous space.
Other works on traditional VMC have studied electronic systems in the TPQ/METTS formalism~\cite{takai2016finite, claes2017finite}, which requires evolving a large number of states, achievable with non-neural approaches.  

\section{Results}

\begin{figure*}[tb]
    \centering
    \includegraphics[width=\textwidth]{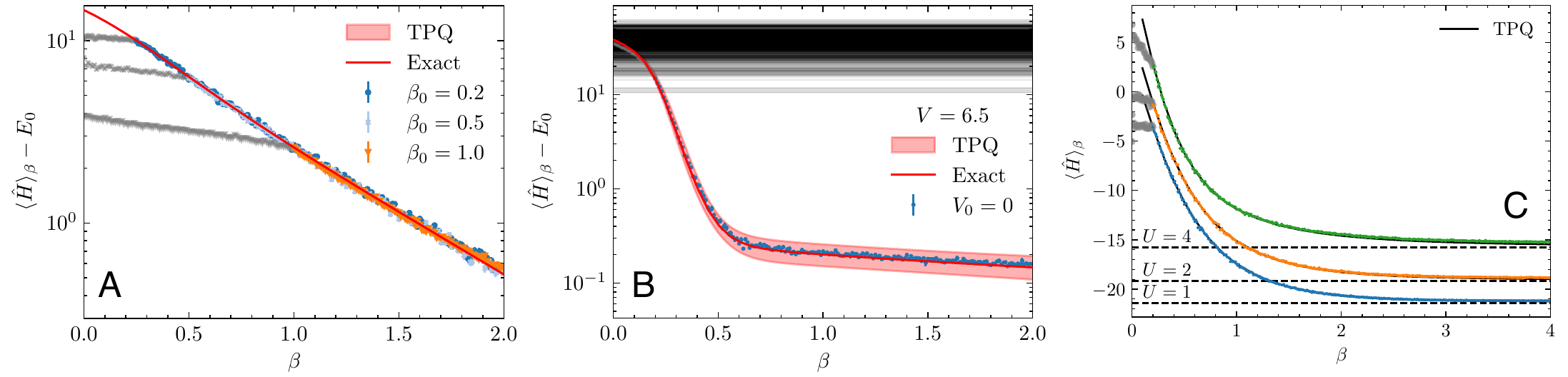}
    \caption{Benchmark results.
    (A) Thermal energies of the \tV model on a $4\times 4$ square lattice at $V=1$ and half-filling, obtained from free fermions ($V_0=0$) at various $\beta_0 = 0.25, 0.5, 1$. Exact thermal energies are shown in red, with $1\sigma$ error bars obtained from TPQ simulations with $1024$ samples. The gray data represents the expectation value of the final Hamiltonian during the work-operator evolution. 
   (B) Thermal energies of the $4 \times 4$ strongly correlated \tV model at doping $\delta=1/8$ and $V = 6.5$, compared to the ED and TPQ predictions. Black lines show the spectrum of $\hat{H}$. (C) Thermal energies of the $4 \times 4$ spinful correlated FH model at doping $\delta=1/8$ at various $U$, compared to the exact TPQ predictions. The ground state is indicated by the dashed horizontal lines.
    }\label{fig:benchmarks}
\end{figure*}

We demonstrate the validity and consistency of the approach for targeting various temperatures on variations of the Hamiltonian
\begin{align}
    \hat{H} &= -t \sum_{\sigma,\langle i,j \rangle} \hat{f}^\dagger_{i,\sigma} \hat{f}_{j,\sigma} + \text{h.c.}  + U \sum_{i} \hat{n}_{i,\uparrow} \hat{n}_{i, \downarrow} + V \sum_{\langle i,j \rangle} \hat{n}_i \hat{n}_j \label{eq:hamiltonian_UV}
\end{align}
where a fully polarized system with $U=0$ yields the \tV model and $V=0$ yields the Fermi-Hubbard model (FH). We set $t=1$ everywhere.

\paragraph*{Small \tV model benchmarks ($U=0$)}
Figure~\ref{fig:benchmarks}(left), shows the results for a thermal state of the \tV model at $V=1$~\cite{romero2025spectroscopy}, prepared from the free-fermion thermal state solution at various inverse temperatures $\beta_0$ (see Appendix~\ref{sec:fhmf} for the mean-field treatment). We observe that the thermal states can be accurately reproduced over a wide range of temperatures. All choices of the reference temperature $\beta_0 \in \{0.25,0.5,1.0\}$ land on the same thermal curve within statistical fluctuations, indicating that the work-operator evolution retains the relevant thermal weight and is effectively path independent for the chosen $\beta_0$. It demonstrates how the work operator can often yield shortcuts to the desired thermal state and validates our adaptive time-stepping approach in tre-pITE.
Furthermore, the approach validates our variational ansatz for representing electronic thermal states of interacting Hamiltonians at various $\beta$. 

Next, we turn to the strongly correlated regime at $V = 6.5$, and doping $\delta = 1/8$ in Fig.~\ref{fig:benchmarks}(center). This doping and coupling regime is relevant for contemporary analog quantum simulation experiments~\cite{chalopin2024probing}, and lies near the onset of the fermionic sign problem's relevance in QMC calculations. In our work, doping improves the Monte Carlo sampling, while the neural representation aids in capturing strongly correlated states. In Fig.~\ref{fig:benchmarks}(middle), we benchmark the method with TPQ and exact diagonalization. We observe that the fNGS tracks the ED/TPQ thermal energies across the entire range of $\beta$, including the high-rank regime where the thermal support spans states well inside the many-body spectrum (black lines), all the way to low-rank states at low temperatures. 

\paragraph*{Doped Fermi Hubbard model ($V=0$)}
We now turn to the doped Fermi-Hubbard model. We first consider a small $4 \times 4$ lattice at $\delta=1/8$ doping at various couplings $U$ in Fig.~\ref{fig:benchmarks}~(right). We observe that our method can predict the thermal state over a wide range of temperatures and eventually converges towards the ground state at high $\beta$ (low $T$). Beyond matching energies, the slope $\partial_\beta \langle \hat{H}\rangle_\beta$ closely follows TPQ, indicating that the heat capacity is captured properly. The small bias that appears only at the largest $\beta$ and $U$ values correlates with the growth of correlations and smaller signals in the fidelity gradients, which is related to the decaying energy variance.
Increasing the hidden dimension of the model systematically removes this bias (data not shown), suggesting that it is a model-capacity effect rather than a limitation of the TFD construction.

\begin{figure*}[tb]
    \centering
    \includegraphics[width=1.0\linewidth]{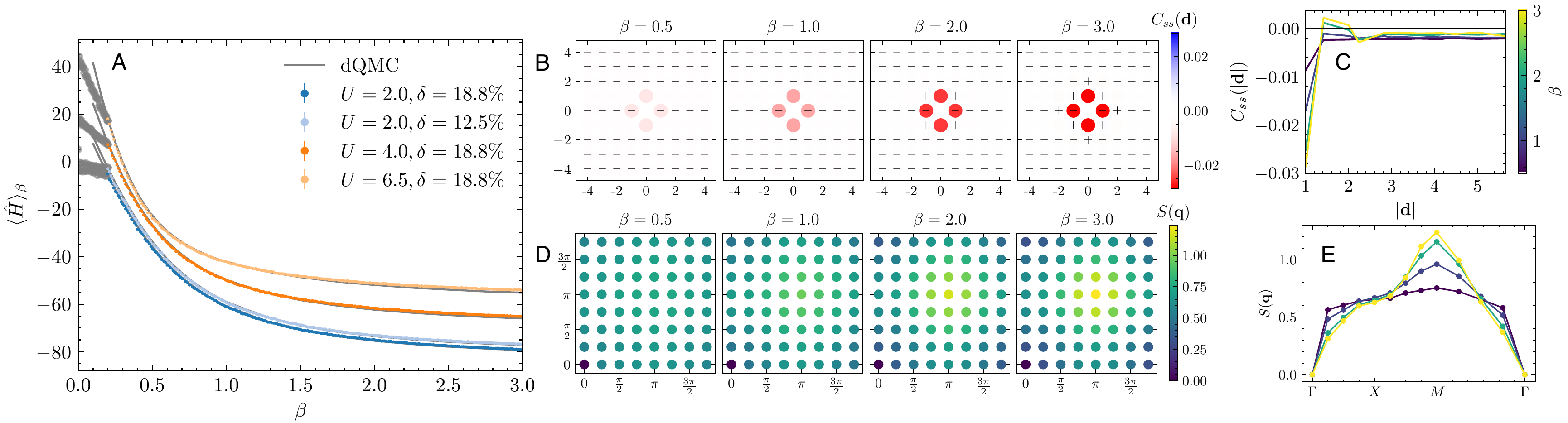}
    \caption{(A) Thermal energies at various couplings and dopings of an $8\times 8$ spinful FH model. For the $U=4$ and $\delta=18.8\%$ case, we show for various temperatures the evolution of the (B,C) connected spin-spin correlation $C_{ss}(\vec{d})$, and (D, E) corresponding spin structure factor $S(\vec{q})$.}
    \label{fig:combined_doping_L8}
\end{figure*}

We now scale our approach to study various dopings on an $8\times8$ torus. 
Panel~A of Fig.~\ref{fig:combined_doping_L8} reports energies for $U\in\{2,4,6.5\}$ and dopings $\delta \in \{12.5\%,18.8\%\}$.
The cooling curves are smooth and monotone, and the $\beta$-dependence separates with interaction and filling.
For the larger interactions ($U=4$ and $6.5$), the onset of stronger curvature at intermediate $\beta$ indicates the growth of antiferromagnetic correlations, which we elaborate on below.
In Fig.~\ref{fig:combined_doping_L8}~(B,C,D,E), we show the equal-time connected
spin-spin correlation function and static spin structure factor,
\begin{align}
C_{ss}(\vec{d})
&= \frac{1}{N}\sum_{i}
   \expval{\hat{S}_i^z \hat{S}_{i+\vec{d}}^z} - \expval{\hat{S}_i^z}\expval{ \hat{S}_{i+\vec{d}}^z} \\
S(\vec q)
&= \sum_{\vec d} e^{i\vec q\cdot \vec d} C_{ss}(\vec d).
\end{align}
where $\vec{d}$ is the displacement vector modulo the lattice dimensions, and $\hat{S}_i^z=\tfrac12(\hat{n}_{i\uparrow}-\hat{n}_{i\downarrow})$. Comparison to TPQ or ED~\cite{weinberg2017quspin} is no longer feasible, and to some extent, differences with dQMC~\cite{SmoQyDQMC} can be attributed to the fact that the latter works in the grand-canonical ensemble, with a remaining variation on the fermion number. We analytically describe differences in correlation observables in Appendix~\ref{sec:infiniteT_expvals}.
For fixed particle numbers one has $S(\vec{0})=0$ and a weak negative tail in $C_{ss}(\abs{\vec{d}}>0)$ at $\beta \to 0$ of order $1/(N-1)$, while grand-canonical averages give $C_{ss}(\vec{d>0}) = 0$ uniformly. We start observing deviations at low temperatures between our predictions at $U=6.5$, which can be attributed to the limited expressiveness of our neural model (see Appendix~\ref{sec:numerical_details} for details). 

Panels~B-E of Fig.~\ref{fig:combined_doping_L8} (shown for $U=4$, $\delta=18.8\%$) visualize the temperature evolution of equal-time spin correlations. At higher temperatures ($\beta \lesssim 1$), $C_{ss}(\vec d)$ is short-ranged and nearly featureless aside from the canonical ensemble offset discussed above. Upon cooling ($\beta \gtrsim 1$), an alternating pattern emerges in real space, and the spin-spin correlations $C_{ss}(\vec{d})$ become increasingly longer range, converging towards an antiferromagnetic state.
This is reflected in $S(\mathbf q)$, which displays enhanced weight at
Néel ordering $\vec{q} = (\pi,\pi)$.

\section{Conclusions and outlook}
We have introduced fermionic neural Gibbs states (fNGS), a variational framework that combines thermofield doubling with neural transformations of geminal pair orbitals to represent correlated fermionic thermal ensembles. Starting from a mean-field reference and evolving with a work operator via tre-pITE, the approach reaches target temperatures efficiently while preserving antisymmetry and scalability. Benchmarks on the doped \tV and Fermi-Hubbard models show that fNGS reproduces thermal energies over wide temperature and interaction ranges, captures spin correlations, and scales beyond exact methods. fNGS emerges as a practical tool for finite-temperature studies of strongly correlated fermions in >1 dimensions, and it charts a clear path to translating modern fermionic neural-quantum-state techniques into a method for understanding thermal phenomena.

Looking ahead, several extensions are natural: (i) extensions to continuum/continuous space approaches where ground-state ansatze are available, see e.g.\ Refs.~\cite {pescia2024message, kim2024ultracold} (ii) response and real-time dynamics from purified fNGS for transport and spectroscopy~\cite{nys2024real} (iii) and larger or more expressive neural models to move into the low-temperature regime relevant to e.g.\ cuprates.

In the context of (iii), it is important to realize that even understanding which types of neural networks and optimization techniques yield accurate ground-state approximations is a very active field of research (see e.g.\ Ref.~\cite{loehr2025enhancingneuralnetworkbackflow, sharma2025comparing, gu2025solving, gao2024neural, liu2024neural, liu2024unifying, moreno2022fermionic, romero2025spectroscopy, pescia2024message} for recent progress and benchmarks), and advancements therein will directly impact the accuracy of the current method as well in the future.
Another possible extension is to consider alternative cooling approaches, such as the exponential cooling approach used in the exponential tensor renormalization group (XTRG) 
~\cite{chen2018exponential}.

\begin{acknowledgments}
J.~Nys would like to thank G.~Carleo and Z.~Denis for insightful discussions in the initial stages of this work, and M.~Medvidović for many useful discussions throughout this work. The simulations were carried out using an extension of NetKet~\cite{carleo2019netket, vicentini2022netket}. TPQ and ED results were produced using QuSpin~\cite{weinberg2017quspin}, and dQMC predictions were produced using the open-source SmoQyDQMC package~\cite{SmoQyDQMC}.
\end{acknowledgments}

\clearpage 
\appendix
\onecolumngrid  

\section{Mean-field Hamiltonians}\label{sec:fhmf}
We consider the Fermi-Hubbard model
\begin{align}
    \hat H = - t \sum_{\sigma\in\{\uparrow, \downarrow\}} \sum_{\langle i, j\rangle} \left[\hat f^\dagger_{i,\sigma} \hat f_{j,\sigma} + \hat f^\dagger_{j, \sigma} \hat f_{i, \sigma}\right] + U \sum_{i}  \hat n_{i, \uparrow} \hat n_{i, \downarrow}
    \label{eq:hamiltonian_fermion}
\end{align}
The mean-field Hamiltonian with respect to a state $\Phi$ with average density $\expval{\hat{n}_{i, \sigma}}_\Phi = \mel{\Phi}{\hat{n}_{i, \sigma}}{\Phi}/\braket{\Phi}{\Phi}$ reads
\begin{align}
    \hat H_0(\Phi) = - t \sum_{\sigma, \langle i, j\rangle} \left[\hat f^\dagger_{i,\sigma} \hat f_{j,\sigma} + \hat f^\dagger_{j, \sigma} \hat f_{i, \sigma}\right] + U \sum_{i}  \hat{n}_{i, \uparrow} \expval{\hat{n}_{i, \downarrow}}_\Phi + \hat{n}_{i, \downarrow} \expval{\hat{n}_{i, \uparrow}}_\Phi + \Esh(\Phi)
\end{align}
The last constant term is $\Esh(\Phi) = - U \sum_{i}  \expval{\hat{n}_{i, \uparrow}}_{\Phi} \expval{\hat{n}_{i, \downarrow}}_{\Phi}$.
Hence, written in terms of an explicit quadratic Hamiltonian
\begin{align}
    \hat{H}_0 &= \sum_{i,j,\sigma} \kappa_{i,j,\sigma}(\Phi) \hat f^\dagger_{i,\sigma} \hat f_{j,\sigma} + \Esh(\Phi) 
\end{align}
we have
\begin{align}
    \kappa_{i, j, \sigma}(\Phi) = -t \delta_{j \in \mathcal{NN}(i)} +   U \expval{\hat{n}_{i, -\sigma}}_\Phi  \delta_{i,j} \label{eq:mf_kappa_fh}
\end{align}

\section{Thermal Hartree-Fock}\label{sec:thf}

The hopping matrix $T$ is defined by $T_{ij}=-t$ on $\langle i,j\rangle$ and $0$ otherwise.
Our thermal Hartree-Fock (THF) ansatz introduces, for each spin $\sigma\in\{\uparrow,\downarrow\}$, two rectangular matrices~\cite{li2025diagonalization}:
\begin{itemize}
    \item $Y^\phi_\sigma\in\mathbb{C}^{N\times N_\sigma}$, whose reduced QR decomposition yields an isometry
$Q^\phi_\sigma$ with ${Q^\phi_\sigma}^{\dagger} Q^\phi_\sigma=\mathbb{I}$, and
    \item $Y^d_\sigma\in\mathbb{R}^{N\times N_\sigma}$, whose reduced QR yields $Q^d_\sigma$ and the site occupations
\begin{equation}
  d_{i\sigma}=\sum_{a=1}^{N_\sigma}\big|(Q^d_\sigma)_{ia}\big|^2,
  \qquad 0\le d_{i\sigma}\le 1,
  \qquad \sum_i d_{i\sigma}=N_\sigma .
\end{equation}
\end{itemize}

The Slater state for spin $\sigma$ is built from the $N_\sigma$ orthonormal columns of $Q^\phi_\sigma$. Its one-body density is
$\rho_\sigma = Q^\phi_\sigma {Q^\phi_\sigma}^{\dagger}$ and
$\bra{\Phi_\sigma} K \ket{\Phi_\sigma}=\mathrm{Tr}\big[{Q^\phi_\sigma}^{\dagger} K Q^\phi_\sigma\big]$ for any one-body matrix $K$. Given the densities $d_{i\sigma}$, the spin-resolved mean-field Hamiltonians are (see Eq.~\eqref{eq:mf_kappa_fh})
\begin{equation}
  K_\uparrow = T + \mathrm{diag}\big(U d_{\downarrow}\big),
  \qquad
  K_\downarrow = T + \mathrm{diag}\big(U d_{\uparrow}\big),
\end{equation}
and the free-energy functional minimized at temperature $T=1/\beta$ is
\begin{equation}
  \mathcal{F}[Q^\phi_\uparrow,Q^\phi_\downarrow,d_\uparrow,d_\downarrow]
  = \sum_{\sigma}\bra{\Phi_\sigma} K_\sigma \ket{\Phi_\sigma}
    - U \sum_i d_{i\uparrow} d_{i\downarrow}
    - \frac{1}{\beta} S[d_\uparrow,d_\downarrow],
\end{equation}
The first term is the band energy of the Slater states in the self-consistent potentials, the second removes the double counting of the on-site interaction (the energy shift in the previous section), and the last accounts for thermal mixing via the occupations $d_{i\sigma}$. We introduced the sitewise Fermi-Dirac entropy
\begin{equation}
  S[d_\uparrow,d_\downarrow]
  = - \sum_{i,\sigma}\big[ d_{i\sigma}\ln d_{i\sigma} + (1-d_{i\sigma})\ln(1-d_{i\sigma})\big].
\end{equation}
Fixing particle numbers $(N_\uparrow,N_\downarrow)$ is built in through the column counts of $Y^\phi_\sigma$ and $Y^d_\sigma$.

Although our free-energy functional uses Fermi-Dirac occupations and an entropy term, i.e.\ it looks grand-canonical, the
Hartree-Fock stationarity equations for the orbitals are identical to the canonical case. The particle-number constraint enters only via a Lagrange multiplier (the chemical potential) that fixes the occupations at the Fermi level. The self-consistent single-particle orbitals $Q^\phi_\sigma$ that diagonalize the mean-field matrices $K_\sigma$ are
unchanged by whether one enforces $N_\sigma$ explicitly (canonical) or through $\mu_\sigma$ (grand-canonical) for the
same target densities. Only the occupation numbers differ by the choice of $\mu$, not the orbitals themselves.

\section{Thermal state as an AGP}\label{sec:pfaffian}
For a mode set $\mathcal{M}$ with $M=|\mathcal{M}|$ and variable particle number $N\in[0,M]$ we introduce the state
\begin{align}
    \ket{A} = \prod_{\mu,\nu\in\mathcal{M}}\left[1 + A_{\mu\nu} \hat{f}_\mu^\dagger \tilde f_\nu^\dagger \right]\ket{\Omega}.
\end{align}
Projecting onto the sector with $N$ particles in each copy~\cite{misawa2019mvmc},
\begin{align}
    \mathcal{P}_{N} \ket{A} &= \ket{\Pf(A)} \\ &= \Bigg[\sum_{\mu,\nu\in\mathcal{M}} A_{\mu\nu} \hat{f}_\mu^\dagger \tilde{f}_\nu^\dagger \Bigg]^{N} \ket{\Omega}.
\end{align}
For a configuration $\ket{x}=\prod_{i=1}^{N} \hat{f}_{p_i}^\dagger \prod_{j=1}^{N} \tilde f_{q_j}^\dagger \ket{\Omega}$, physical and auxiliary modes never overlap
\begin{align}
    \braket{x}{\Pf(A)} = \det\big[A_{p_i,q_j}\big]_{i,j=1}^{N},
\end{align}
and the Pfaffian reduces to a determinant over the pair matrix subblock $A_{p,q}$.

\section{Variational model}\label{sec:model}
We start by encoding the single and relative site properties. The occupation per site $o_i \in \{0, \uparrow, \downarrow, \uparrow \downarrow\}$ is encoded through a parametrized vector $E_O(o_i) \in \mathbb{R}^{d_h}$, and similarly the position is encoded with an embedding vector $E_R(\vec{r}_i) \in \mathbb{R}^{d_h}$ with hidden dimension $d_h$. We encode the (minimal distance) positive-elements distance vector $\vec{a}_{ij} = \abs{\vec{r}_i - \vec{r}_j}$ and the norm $r_{ij} = \norm{\vec{r}_i - \vec{r}_j}$ through embeddings $E_{A}(\vec{a}_{ij}) \in \mathbb{R}^{d_h}$ and $E_{D}(\vec{r}_{ij}) \in \mathbb{R}^{d_h}$, together yielding the encodings
\begin{align}
    v_i &\leftarrow \textrm{LayerNorm}(E_O(o_i) + E_R(\vec{r}_i)) \in \mathbb{R}^{d_h} \\
    e_{ij} &\leftarrow \textrm{LayerNorm}(E_{A}(\vec{a}_{ij}) + E_{D}(\vec{r}_{ij})) \in \mathbb{R}^{d_h}
\end{align}
with $d_h$ the hidden dimension and where we add the encodings element-wise. We use the same approach for the auxiliary sites $\tilde{v}_i$, using the same embeddings. 

Without backflow, we would take these encodings and construct the pair-orbital input
\begin{align}
    u_i &= W \cdot \textrm{LayerNorm}(v_i) 
\end{align}
where $W \in \mathbb{R}^{d_h \times d_h}$ (same for the auxiliary system).
We then combine
\begin{align}
    l_{ij} &= \textrm{LayerNorm}(\left[u_i + \tilde{u}_j, u_i \odot \tilde{u}_j, \abs{u_i - \tilde{u}_j}, E_{A}'(\vec{a}_{ij})\right]) \\
    l_{ij} &\leftarrow \textrm{FFN}(l_{ij}) \\
    m_{ij} &= [l_{ij}, W \cdot e_{ij}]
\end{align}
where square brackets indicate concatenation. This layer allows for a flexible and expressive representation of relative site properties.
We then evaluate the pair orbitals for a pair of electrons at physical mode $(i, \sigma)$ and unphysical mode $(j, \tau)$ as
\begin{align}
    [\varphi(x, \tilde{y})]_{(i,\sigma), (j, \tau)} &= \textrm{MLP}_{\sigma, \tau}(m_{ij}) e^{-\alpha r_{ij}^2} 
\end{align}
with an MLP last RBM-like layer of complex output with $d_h$ hidden nodes of which we take the product after applying the activation function $\cosh$. All other activation functions will be taken as GELU.

With backflow, we first transform the node and edge properties before applying the transformations above. We do this by first using a constant key-query reminiscent of the Vision Transformer (as described in Refs.~\cite{viteritti2023transformer, rende2025queries}) to gather intra-species correlations
\begin{align}
    v_i &= \textrm{ViT}(v) \\
    \tilde{v}_j &= \textrm{ViT}(\tilde{v})
\end{align}
where $v$ contains all site information per species. We use the same ViT for both the physical and auxiliary species. Our ViT implementation does not consider patching, which should be included in future work to improve model expressivity.

Next follows an inter-species correlator using bi-partite attention. Therefore, we introduce an architecture called BiViT. Its overall goal is to learn edge embeddings between physical-auxiliary pair sites, i.e.\
\begin{align}
    e_{ij} &\leftarrow \textrm{BiViT}(v, \tilde{v}, e)
\end{align}
The BiViT operates in detail as follows. Assume an input $v_i$ and $\tilde{v}_i$ coming from the intra-species ViT, we have
\begin{align}
    \tilde{u}_i^H &= W_I^H \cdot \textrm{LayerNorm}(\tilde{v}_i) \\
    u_i^H &= \sum_j \alpha_{ij}^H \tilde{u}_j^H \\
    u_i &\leftarrow v_i + W_O \cdot [W_C^H \cdot u_i^H]_{H = 1, ..., N_H} \\
    u_i &\leftarrow v_i + \textrm{FFN}(\textrm{LayerNorm}(u_i))
\end{align}
for one of the $N_H$ heads $H$, and $W_I^H\in \mathbb{R}^{d_{eff} \times d_h}$, $W_C^H\in \mathbb{R}^{d_{eff} \times d_{eff}}$, $W_O^H\in \mathbb{R}^{d_h \times d_h}$, where $d_{eff} = d_h / N_H$. Furthermore, $\alpha_{ij}^H \in \mathbb{R}$ is a trainable mask with translation symmetry imposed, i.e.\ we parametrize $\alpha_{0j}^H$ and obtain other entries through translation. We symmetrically use the same transformation for the other system with the same parameters. The following sequence 
\begin{align}
    u &\leftarrow \textrm{ViT}(v) \\
    \tilde{u} &\leftarrow \textrm{ViT}(\tilde{v}) \\
    \varepsilon &\leftarrow \textrm{BiViT}(u, \tilde{u}, e)
\end{align}
is then iterated $N_L$ times (typically $N_L = 2$) using different parameters per layer.
Before plugging the result into the post-backflow transformations above, we set
\begin{align}
    v_i &\leftarrow v_i + u_i \\
    \tilde{v}_j &\leftarrow \tilde{v}_j + \tilde{u}_j \\
    e_{ij} &\leftarrow e_{ij} + \varepsilon_{ij} 
\end{align}
In short, the node encodings gather the intra-system correlations, while the edge encodings gather the inter-correlations. It is straightforward to verify that this model exactly contains the infinite temperature state, and reduces to a pure-state neural network approximation in the low-$T$ limit.
Irrespective of the details of the architecture, we emphasize that it is important to capture the interactions between the two systems (in our case, through BiViT), by encoding the bipartite (inter-system) edges $e_{ij}$. Without this block, or when the expressiveness in this block is limited, simulations are expected to yield (much) less accurate results.

We further include two types of bosonic correlation factors: inter- and intra-body Jastrow factors. For the intra-system correlations, we use the standard two-body form for both systems
\begin{align}
J(x) &= \sum_{i,j} W^{\sigma_i, \sigma_j}_{i, j} n_{i,\sigma_i} n_{j,\sigma_j}     
\end{align}
with spin-orbital occupation numbers $n_{i, \sigma}$.
For the inter-system correlations, we encode the site occupations $o_i$ through a one-hot encoding, and form the two-site Jastrow
\begin{align}
    J(x, \tilde{y}) &= \sum_{i, j} W_{i,j}^{k} [\textrm{one-hot}([n_{i, \uparrow}, n_{i, \downarrow}])]_k  [\textrm{one-hot}([\tilde{n}_{j, \uparrow}, \tilde{n}_{j, \downarrow}])]_k
\end{align}

\section{Observables}\label{sec:observables}
Expectation values follow from
\begin{align}
  \expval{\hat O}_\beta
  = \frac{\mel{\Psi(\beta)}{\hat O \otimes \tilde{\iden}}{\Psi(\beta)}}{\braket{\Psi(\beta)}} .
\end{align}
We estimate this quantity with Monte Carlo sampling in the doubled configuration space. Let $x$ and $\tilde y$ denote physical and auxiliary configurations in the computational basis. Define
\begin{align}
  \pi(x,\tilde y) &= \frac{\abs{\Psi(x,\tilde y)}^2}{\sum_{x',\tilde y'} \abs{\Psi(x',\tilde y')}^2} , \\
  O_{\mathrm{loc}}(x,\tilde y) &= \frac{\mel{x,\tilde y}{\hat O \otimes \tilde{\iden}}{\Psi}}{\Psi(x,\tilde y)}
  = \sum_{x'} \mel{x}{\hat O }{x'} \frac{\Psi(x',\tilde y)}{\Psi(x,\tilde y)} .
\end{align}
Then
\begin{align}
  \expval{\hat O}_\beta = \mcest{(x,\tilde y)\sim \pi}{O_{\mathrm{loc}}(x,\tilde y)} .
\end{align}
The Monte Carlo samples are produced using the sampler proposed in Appendix~\ref{sec:sampling}.

\section{Monte Carlo sampling in a doubled space}\label{sec:sampling}
Sampling is performed using the Metropolis-Hastings algorithm, employing a composite transition kernel comprising multiple proposal rules. The first rule, termed `site exchange,' facilitates nearest-neighbor transitions. This includes: (i) single-electron hopping, (ii) spin exchange between two electrons, and (iii) doublon hopping. In the doubled space, this rule is applied by randomly selecting either the physical or the auxiliary system for the update.

The second rule, `diagonal exchange,' is particularly useful at high temperatures. It executes symmetric single-electron moves simultaneously on both the physical and auxiliary systems. Therefore, we restrict proposals to moves that are valid for both configurations at the same time. 

Finally, to restore symmetry between the two systems, we employ a global `swap rule' that attempts to exchange the physical and auxiliary configurations entirely.

The above rules are respectively performed with a 79\%, 20\%, and 1\% probability rate.

\section{Alternative ansatz}\label{sec:alternative_ansatz}
For completeness, we introduce an alternative ansatz based on considering the physical+auxiliary system in its entirety as an interacting fermionic system with $2N$ electrons. If we consider the TFD system, a generic assumption for the wave function would be
\begin{align}
    \Psi(x, \tilde{y}) &= \sum_{\mathcal{S}} w_\mathcal{S}(x, \tilde{y}) \det{\Phi(x, \tilde{y})}_{\mu, k} \label{eq:psi_double_orbs}
\end{align}
where we bundle the physical and unphysical coordinates into a larger index $k=(p, \tilde{q})$. This represents an antisymmetric wave function on the doubled system, with orbitals spanning both systems. Since the orbitals are expected to be the same on both systems, i.e.\ the orbitals $\mu$ come in pairs, we can perform a decomposition, where the set $\mathcal{S}$ splits up into two sets. This procedure is similar to defining spin-orbitals, and taking into account that physical and unphysical fermions cannot occupy each other's modes. 
\begin{align}
    \Psi(x, \tilde{y}) &= \sum_{\mathcal{S}} w_\mathcal{S}(x, \tilde{y}) \det{\Phi(x, \tilde{y})}_{\mu, p} \det{\Phi^*(\tilde{y}, x)}_{\mu, q}
\end{align}

If we further assume that $w_\mathcal{S}(x, \tilde{y}) = \prod_{\mu}^{N} w_{\mu}(x, \tilde{y})$ we can connect this ansatz to ours
\begin{align}
    \Psi(x, \tilde{y}) &= \det{\sum_\mu w_\mu(x, \tilde{y}) [\Phi(x, \tilde{y})]_{\mu, p} [\Phi(\tilde{y}, x)]_{\mu, q}^*} \label{eq:psi_double_orbs_combined}
\end{align}
which resembles a Bogoliubov-de Gennes determinant.
Note that in the above $w$ is sample dependent.
This can be cast into the form
\begin{align}
    \Psi(x, \tilde{y}) &= \det{\varphi(x, \tilde{y})}_{p q}
\end{align}
where $\varphi$ is Hermitian and
\begin{align}
    [\varphi(x, \tilde{y})] &= [\Phi(x, \tilde{y})]^T \textrm{diag} [w(x, \tilde{y})] [\Phi(\tilde{y}, x)]^*
\end{align}
Note that the form in Eq.~\eqref{eq:psi_double_orbs_combined} reduces as follows to the mean-field pair orbitals
\begin{align}
    [\Phi(x, \tilde{y})]_{\mu, p} \to [\Phi]_{\mu, p}, \quad w_\mu(x, \tilde{y}) \to e^{-\beta \epsilon_\mu/2}
\end{align}
When approaching a ground state, we have
\begin{align}
    \det{\sum_\mu [\Phi(x)]_{\mu, p} [\Phi(\tilde{y})]_{\mu, q}^*} 
\end{align}
Or, in other words, the dependency of the orbitals on the other system vanishes. At infinite temperatures $\beta \to 0$, we get
\begin{align}
    \det{\sum_\mu \delta_{\mu, p} \delta_{\mu, q}}  &\propto \det{ \delta_{p,q} }
\end{align}
or any permutation between the matching of $\mu$ and $p$.
This illustrative discussion demonstrates how the low- and high-temperature constraints are automatically built into our ansatz, since the latter forms a special case of our more general ansatz used in this work.

\section{Imaginary time evolution\label{sec:adaptive_trepITE}}

\subsection{Projections}
We use a second-order consistent Taylor-root expansion of the imaginary-time propagator, as introduced in Ref.~\cite{nys2024ab}.
We approximate the target state $\hat V\ket{\phi}$ by maximizing the fidelity of a trial state $\ket{\psi}$ with respect to it,
\begin{align}
  \mathcal{F}\left[\ket{\psi},\hat V\ket{\phi}\right]
  &= \frac{\mel{\psi}{\hat V}{\phi}}{\braket{\psi}}
     \frac{\mel{\phi}{\hat V^\dagger}{\psi}}{\mel{\phi}{\hat V^\dagger \hat V}{\phi}} .
  \label{eq:fidelity}
\end{align}
Monte Carlo estimators for this objective follow the constructions in Ref.~\cite{medvidovic2021classical}, with the variance-reduced, control-variate form of Ref.~\cite{sinibaldi2023unbiasing} (see also Ref.~\cite{gravina2024neural} for a pedagogical and systematic review on fidelity estimators and their gradients):
\begin{align}
    F &= \mcest{x \sim \abs{\psi}^2}{f_1(x)} F_2 + F_{CV} ,\\
    F_{CV} &= -\tfrac{1}{2}\left(\mcest{x \sim \abs{\psi}^2}{g_1(x)} G_2 - 1\right) ,
\end{align}
where
\begin{align}
    f_1(x) &= \frac{[\hat V \phi](x)}{\psi(x)}, &
    f_2(y) &= \frac{\psi(y)}{[\hat V \phi](y)}, &
    g_i(x) &= \abs{f_i(x)}^2, \\
    F_2 &= \mcest{y \sim \abs{\hat V \phi}^2}{f_2(y)}, &
    G_2 &= \mcest{y \sim \abs{\hat V \phi}^2}{g_2(y)} .
\end{align}
Rao-Blackwellization is used to further cut variance, as in Refs.~\cite{medvidovic2021classical,gravina2024neural}. The control-variate contribution is biased for finite samples, but its bias vanishes as $F \to 1$. We used the short-hand notation
\begin{align}
    [\hat V \phi](x) = \mel{x}{\hat V}{\phi} = \sum_{x'} \mel{x}{\hat V}{x'} \phi(x') .
\end{align}

To eliminate sampling from $\hat V\ket{\phi}$, we adopt self-normalized importance sampling using configurations from $\ket{\phi}$~\cite{nys2024real, nys2024ab}. For any $f(y)$ that stands for $f_2$ or $g_2$,
\begin{align}
  \mcest{y \sim \abs{\hat V \phi}^2}{f(y)}
  \;\to\;
  \frac{\mcest{y \sim \abs{\phi}^2}{w(y) f(y)}}
       {\mcest{y \sim \abs{\phi}^2}{w(y)}},\qquad
  w(y) = \abs{\tfrac{[\hat V \phi](y)}{\phi(y)}}^2 .
\end{align}
Near convergence one may instead draw $y \sim \abs{\psi}^2$ since $\ket{\psi}\approx \hat V\ket{\phi}$, yielding
\begin{align}
  \mcest{y \sim \abs{\hat V \phi}^2}{f(y)}
  \;\to\;
  \frac{\mcest{y \sim \abs{\psi}^2}{w(y) f(y)}}
       {\mcest{y \sim \abs{\psi}^2}{w(y)}},\qquad
  w(y) = \abs{\tfrac{[\hat V \phi](y)}{\psi(y)}}^2 \approx 1 .
\end{align}
However, we opt to sample from $\abs{\phi}^2$ instead, since we will see that the gradients have $F_2$ included as an overall magnitude factor, and can therefore be computed every e.g.\ $n=5$'th step to reduce computational cost.

When restricting $\ket{\psi}$ to a variational family $\ket{\psi(\theta)}$, gradients of \eqref{eq:fidelity} can be estimated as in Ref.~\cite{medvidovic2021classical}:
\begin{align}
  \partial_{\theta_k} F
  &= \mcest{x \sim \abs{\psi}^2}{ O_k(x)^* \left(f_1(x) - F_1\right)} F_2
   = \mcest{x \sim \abs{\psi}^2}{ \Delta O_k(x)^* f_1(x) F_2 } ,
\end{align}
with $O_k(x)=\partial_{\theta_k}\log\psi(x)$ and
\begin{align}
  \Delta O_k(x) = O_k(x) - \mcest{x' \sim \abs{\psi}^2}{O_k(x')} .
\end{align}
Since $F_2$ is a global factor, the update direction is unaffected by it. We therefore prefer the asymmetric fidelity in \eqref{eq:fidelity}, where the operator acts only on $\ket\phi$, avoiding additional importance sampling from $\ket\psi$ required by symmetric or Padé-type schemes~\cite{gravina2024neural} that can introduce (finite sample) biases in the gradient direction. This becomes particularly important for fermionic systems where many small contributions and zeros occur.
An alternative, CV-aware gradient form is
\begin{align}
  \partial_{\theta_k} F
  = \mcest{x \sim \abs{\psi}^2}{ O_k(x)^* \left(f_1(x) F_2 - F\right)} ,
\end{align}
where $F$ denotes the control-variate adjusted fidelity.

The quantum geometric tensor is estimated by
\begin{align}
  S_{kk'} = \mcest{x \sim \abs{\psi}^2}{ \Delta O_k(x)^* \Delta O_{k'}(x) } .
\end{align}

If gradient estimates become unstable, for example, when
\begin{align}
  \abs{O_k(x)} \gg \abs{\psi(x)} ,
\end{align}
we introduce an auxiliary sampling density $q(x)$ and reweight all expectations over $x$ using self-normalized importance sampling,
\begin{align}
  \mcest{x \sim \abs{\psi}^2}{f(x)}
  \;\to\;
  \frac{\mcest{x \sim q}{w(x) f(x)}}
       {\mcest{x \sim q}{w(x)}},\qquad
  w(x)=\frac{\abs{\psi(x)}^2}{q(x)} ,
\end{align}
with the same weights applied to the metric $S_{kk'}$. In particular, for small systems and/or high interactions, we find it useful to sample with umbrella sampling 
\begin{align}
    q(x) &\propto e^{-\alpha V(x)} \\
    V(x) &= \log \abs{\Psi(x)}
\end{align}
where the standard form is $\alpha=2$ corresponding to sampling from the Born distribution. This improves the gradient estimation when the wave function becomes peaked or fermions become localized. In an ideal scenario, $\alpha$ can be adjusted dynamically~\cite{misery2025looking}, but yields a prohibitive overhead in the large-scale simulations presented here.

\subsection{Adaptive evolution}
We introduce a cheap adaptive form of tre-tVMC (which in the imaginary-time domain we refer to as tre-pITE, in accordance with Ref.~\cite{nys2024real}, referring to Taylor-root expansions projected imaginary-time evolution) that increases the time step $\dtau$ to render the evolution more efficient. As an efficient approach, we estimate the infidelity of the time-evolved state, i.e.\ $e^{-\tau \hat{H}}\ket{\phi}$, and the unoptimized state $\ket{\phi}$, as a measure for the change expected in the Hilbert space.  We estimate the leading contribution to the infidelity between $e^{-\dtau \hat{H}}\ket{\phi}$ and $\ket{\phi}$ as
\begin{align}
    \mathcal{I}(\dtau) &= \frac{\abs{\mel{\phi}{e^{-\dtau \hat{H}}}{\phi}}^2}{\norm{\ket{\phi}}^2 \norm{e^{-\dtau \hat{H}}\ket{\phi}}^2} \\
    &= \dtau^2 \Var{\hat{H}},
\end{align}
and choose $\dtau = \sqrt{I_c/\Var{\hat{H}}}$. Here $\hat{H}$ denotes either the work operator or the physical Hamiltonian, depending on the step.
We defined a cutoff $I_c$ of maximum initial infidelity. This additionally implies that $\dt \propto N^{-1}$~\cite{wu2023variational}. Notice that $\hat{H}$ can represent the work operator or simply the Hamiltonian applied to the physical part. This scheme is particularly powerful for small systems and is currently only used for those examples.

\subsection{Fast convergence through parameter dynamics}
A second significant improvement to the method is to approximate the dynamics in parameter space throughout the evolution. Said differently, we estimate the velocity $\theta_t$ from the parameter solutions $\theta_t$ and $\theta_{t+\dt_1}$, after evolving from $t$ to $t+\dt_1$. We use this to obtain an initial guess of the solution $\theta_{t+\dt_1 + \dt_2}$ at time $t+\dt_1 + \dt_2$, when evolving over a time $\dt_2$
\begin{align}
    \dot{\theta}_t &= (\theta_{t + \dt_1} - \theta_{t}) / \dt_1 \\
    \theta_{t+\dt_1+\dt_2} &= \theta_{t+\dt_1} + \dot{\theta}_t \dt_2
\end{align}
In practice, this often results in significant speedups, reaching faster convergence.

\section{Analytic expressions for infinite-temperature correlators}\label{sec:infiniteT_expvals}

We consider spinful fermions on $N=L^2$ sites with double occupancy allowed.
At infinite temperature the Hamiltonian becomes irrelevant, and expectation
values reduce to combinatorial averages. We treat the canonical and
grand-canonical ensembles in turn. Throughout we denote spin-resolved
filling fractions by
\begin{align}
    p_\sigma \equiv \frac{N_\sigma}{N}, \qquad
    n \equiv p_\uparrow + p_\downarrow,
\end{align}

\subsection{Canonical ensemble}
In the canonical ensemble (considered in this work) $N_\uparrow$ and $N_\downarrow$ are fixed. The number
of states is $\binom{N}{N_\uparrow}\binom{N}{N_\downarrow}$. At infinite
temperature the spin-up and spin-down configurations are independent.

We obtain the single-site expectations:
\begin{align}
    \langle \hat{n}_{i\sigma}\rangle &= p_\sigma, \\
    \langle \hat{n}_{i\uparrow} \hat{n}_{i\downarrow}\rangle &= p_\uparrow p_\downarrow .
\end{align}
For $i\neq j$:
\begin{align}
    \langle \hat{n}_{i\sigma}\hat{n}_{j\sigma}\rangle
        &= \frac{N_\sigma(N_\sigma-1)}{N(N-1)}, \\
    \langle \hat{n}_{i\uparrow}\hat{n}_{j\downarrow}\rangle
        &= p_\uparrow p_\downarrow .
\end{align}

Since $\langle \hat{S}_i^z\rangle=\tfrac12(p_\uparrow-p_\downarrow)$, the connected on-site variance is
\begin{align}
    C_{ss}(0) &=\langle (\hat{S}_i^z)^2\rangle_c
    = \tfrac14 \left[p_\uparrow(1-p_\uparrow)+p_\downarrow(1-p_\downarrow)\right]
\end{align}
where $\expval{A B}_c = \expval{A B} - \expval{A}\expval{B}$.
For $i\neq j$ the connected correlator is
\begin{align}
    C_{ss}(\vec{d} \neq 0) &= \langle \hat{S}_i^z \hat{S}_{j}^z\rangle_c \\
    &= \tfrac14\!\left(
        \tfrac{N_\uparrow(N_\uparrow-1)}{N(N-1)}
      + \tfrac{N_\downarrow(N_\downarrow-1)}{N(N-1)}
      - 2p_\uparrow p_\downarrow \right)
      - \tfrac14(p_\uparrow-p_\downarrow)^2 \nonumber\\
    &= -\frac{p_\uparrow(1-p_\uparrow)+p_\downarrow(1-p_\downarrow)}{4(N-1)} \\
    &= -\frac{C_{ss}(\vec{0})}{(N-1)}.
\end{align}
Hence, $C_{ss}(d>0) \neq 0$.

The spin structure factor at $\vec{q} = 0$ reads
\begin{align}
    S(\vec{0}) &= \sum_{\vec{d}} C_{ss}(\vec{d}) \\
    &= C_{ss}(0) + (N-1)C_{ss}(\vec{d} \neq 0) \\
    &= 0
\end{align}

For $8\times8$, $N_\uparrow=N_\downarrow=26$, $N=64$ and $p_\uparrow=p_\downarrow \equiv p=26/64$, hence
\begin{align}
    C_{ss}(0)=\langle (\hat{S}_i^z)^2\rangle_c
      &= \frac{p(1-p)}{2}
       \approx 0.12061,\\
    C_{ss}(d>0)=\langle \hat{S}_i^z \hat{S}_{i+\mathbf d}^z\rangle_c
      &= -\frac{p(1-p)}{2(N-1)}
       \approx -0.0019144.
\end{align}

\subsection{Grand-canonical ensemble}

In (determinant) QMC, the system is typically sampled in the grand-canonical
ensemble, where $N_\uparrow$ and $N_\downarrow$ fluctuate. At infinite
temperature the Boltzmann weight is independent of $t$ and $U$, and observables
reduce to averages over independent occupations determined by the fugacities
$e^{\beta\mu_\sigma}$.

For spin $\sigma$ the single-site partition function is
\begin{align}
    z_\sigma = 1+e^{\beta\mu_\sigma},
\end{align}
and the average filling fraction is
\begin{align}
    p_\sigma = \langle n_{i\sigma}\rangle
             = \frac{1}{1+e^{-\beta\mu_\sigma}}.
\end{align}
To impose a desired $p_\sigma$ choose
\begin{align}
    \mu_\sigma(\beta) = \tfrac{1}{\beta}\log\!\left(\tfrac{p_\sigma}{1-p_\sigma}\right),
\end{align}
and for spin balance with total density $n=p_\uparrow+p_\downarrow$,
\begin{align}
    \mu_\uparrow=\mu_\downarrow
      = \tfrac{1}{\beta}\log\!\left(\tfrac{n/2}{1-n/2}\right).
\end{align}

In the $\beta\to 0$ limit, sites and spins are uncorrelated, hence
\begin{align}
    \langle (\hat{S}_i^z)^2\rangle_c
        &= \tfrac14 \left[p_\uparrow(1-p_\uparrow)+p_\downarrow(1-p_\downarrow)\right], \\
    \langle \hat{S}_i^z \hat{S}_{j\neq i}^z\rangle_c &= 0,
\end{align}
\begin{align}
    C_{ss}(\vec d)\equiv \big\langle \hat S_i^z \hat S_{i+\vec d}^z\big\rangle_c
=
\begin{cases}
\tfrac14\left[p_\uparrow(1-p_\uparrow)+p_\downarrow(1-p_\downarrow)\right] , & \vec d=\vec 0,\\[6pt]
0, & \vec d\neq \vec 0,
\end{cases}
\end{align}
Notice the difference compared to the canonical case for $\vec{d} \neq 0$.

This difference now yields a spin structure factor that is flat:
\begin{align}
    S(\mathbf q)
    = \tfrac14\left[p_\uparrow(1-p_\uparrow)+p_\downarrow(1-p_\downarrow)\right].
\end{align}
and hence does not vanish at $\vec{q}=\vec{0}$.

For $8\times8$ at $\langle \hat{N}_\uparrow\rangle=\langle \hat{N}_\downarrow\rangle=26$, where $\hat{N}_\sigma = \sum_i \hat{n}_{i,\sigma}$.
With $p=26/64$,
\begin{align}
    C_{ss}(0)=\langle (\hat{S}_i^z)^2\rangle_c
      &= \frac{p(1-p)}{2}
       \approx 0.12061,\\
    C_{ss}(d>0)
      &= 0.
\end{align}

\paragraph*{Summary} At $T=\infty$ in the \emph{canonical} ensemble, $N_\uparrow$ and $N_\downarrow$ are fixed, so sites are sampled without replacement. This induces weak negative correlations between distinct sites:
\begin{align}
    C_{ss}(\vec d\neq 0)=-\frac{C_{ss}(0)}{N-1}\neq 0
\end{align}
Consequently, the connected structure factor at zero momentum vanishes.
Equivalently 
\begin{align}
    S(\vec 0)=\frac{1}{N} \mathrm{Var}\left(\sum_i \hat S_i^z\right)=0
\end{align}
because the total magnetization is fixed. 
In the \emph{grand-canonical} ensemble, site occupations are independent at $T=\infty$, hence $C_{ss}(\vec d\neq 0)=0$ and the connected $S(\vec q)$ is flat.

\section{Numerical details}\label{sec:numerical_details}
We use a (importance-sampling) weighted version of the minSR and SPRING optimizers~\cite{chen2024empowering, goldshlager2024kaczmarz}. During the fidelity optimization, we use an exponential decaying regularization $\lambda = 10^{-2} \to 10^{-6}$. The SPRING momentum is set to $\mu=0.95$ throughout. The base time step is $\delta \tau = 0.01$. We use a second-order-consistent ($K=2$) Taylor-root expansion for the time propagator, requiring one additional intermediate compression step~\cite{nys2024ab}. In the first step, we use the mean-field target state and optimize over $1500$ steps, starting from $\lambda = 10^{-1}$. All other fidelity optimizations consist of $125$ steps.
In all simulations, we use $1024$ samples and an equal number of Markov chains. Our model consists of $2$ BiViT layers with $d_h = 24$ and $6$ heads. For an $8 \times 8$ system, this results in $\pm 50$k parameters. All simulations are run on a single NVIDIA GH200 GPU.

\bibliography{biblio}

\end{document}